\documentclass[aps,nofootinbib,checkin,prc,preprintnumbers,amsmath,amssymb]{revtex4}

\usepackage{graphicx}
\usepackage{dcolumn}
\usepackage{bm}
\usepackage{epstopdf}

\makeatletter
\AtBeginDocument{\let\($\let\)$
\@ifpackageloaded{natbib}{\ifNAT@numbers\if@filesw\immediate\write\@auxout{\string\global\string\NAT@numberstrue}\fi\fi}{}}
\makeatother
\begin{document}

\def\au#1{{#1}}
\def\f#1{{#1}}
\def\s#1{{#1}}
\def\at#1{{#1}}
\def\pt#1{{#1}}
\def\bt#1{{#1}}
\def\v#1{{\bf #1}}
\def\pg#1{{#1}}
\def\yr#1{{#1}}
\def\pbl#1{{#1}}
\def\cny#1{{#1}}
\def\cty#1{{#1}}
\def\bibitemo#1{\bibitem{#1}}
%\lrh{YACINE MEHTAR-TANI AND GEORG WOLSCHIN}

%\rrh{BARYON STOPPING AND SATURATION PHYSICS IN...}

\title{Baryon stopping and saturation physics in relativistic collisions}
%\title{Baryon stopping and saturation physics in relativistic collisions}
\author{Yacine Mehtar-Tani}
\author{Georg Wolschin}
\affiliation{Institut f{\"ur} Theoretische Physik der Universit{\"a}t Heidelberg, Philosophenweg 16, D-69120 Heidelberg, Germany}

%\begin{history}
%\re{30 July 2009}
%\end{history}

\begin{abstract}
We investigate baryon transport in relativistic heavy-ion
collisions at energies reached at the CERN Super Proton Synchrotron (SPS),
BNL Relativistic Heavy-Ion Collider (RHIC), and CERN's Large Hadron Collider (LHC)
in the model of saturation.
An analytical scaling law is derived within the color glass condensate
framework based on small-coupling QCD. Transverse momentum
spectra, net-baryon rapidity distributions, and their energy, mass, and
centrality dependencies are well described.
In  comparison with RHIC data
in Au + Au collisions at
\(\sqrt {s_{NN}}\) = 62.4  and 200 GeV, the gradual approach to the
gluon saturation regime is investigated and limits for the saturation-scale exponent are determined.
Predictions for net-baryon rapidity spectra and the mean rapidity loss in
central Pb + Pb collisions at LHC energies of \(\sqrt {s_{NN}}\) = 5.52 TeV are made.
\end{abstract}

%\pacs{24.85.+p, 25.75.-q, 25.75.Dw, 12.38.Mh}
%\keywords Relativistic heavy-ion collisions; baryon stopping; color glass condensate; mass-number dependence
\maketitle

\section{\label{sec:intro}Introduction}
Gluon saturation has been the focal point of important and interesting particle-physics investigations for many years. Its observation would allow to  access a new regime of quantum chromodynamics where high-density gluons form a coherent state. In regions of large parton densities the physics is governed by a single hard scale, \(Q_s \gg \Lambda_{\text{QCD}}\), which increases with energy and thus allows the use of small-coupling techniques \cite{mcl94}. In this regime, gluon recombination starts to compete with the exponentially increasing gluon splitting and the gluon distribution function is expected to saturate.
%such that unitarity is preserved.

At the HERA, some evidence for gluon saturation in the proton was found in deep inelastic \(e + p\) collisions at high energy and low values of Bjorken-\(x\), but the results are still open to interpretation \cite{sim06}. The existence of geometric scaling as predicted by the color glass theory as an approach to saturation physics was indeed  confirmed, constituting the most important evidence for saturation so far \cite{sta01}.

Since the saturation scale is enhanced by a factor  \(A^{1/3}\) in heavy ions, as compared to protons, it is natural to investigate saturation in relativistic heavy-ion collisions, as has been done by many authors \cite{arm09}. Here theoretical  QCD-based approaches have, so far, usually concentrated on charged-hadron production and, in the central rapidity region, a reasonable understanding was achieved in the color glass condensate framework \cite{mcl94,Bal96,Jal97,Ian01} through inclusive gluon production \cite{nar05,alb07bis}.

In \(p + A\) collisions the nuclear wave function is being probed at small \(x\) in the forward direction and it should show saturation below a characteristic value of \(x\). Hence, the effect of gluon condensation and quantum evolution should be measurable, as was attempted by the BRAHMS Collaboration at the Relativistic Heavy-Ion Collider (RHIC) in \(d\) + Au collisions \cite{ars04}. Measurements of the nuclear modification factor showed the Cronin enhancement at midrapidity, but the modification factor is suppressed at forward rapidities.  This is in qualitative agreement with the color-glass predictions \cite{kha03,kha04}.

Experimental heavy-ion investigations at the Large Hadron Collider (LHC)  concentrate on the midrapidity region since \textsc{alice} \cite{aam08} covers rapidities up to \(|y|\) = 2. It  provides measurements of lower \(x\) values than before---down to \(10^{-5}\)---at an energy scale that is high enough to provide crucial tests of gluon saturation.

In the present phenomenological investigation we use the transverse momentum spectra and rapidity distributions of net baryons (\(B-\bar{B}\)) in relativistic heavy-ion collisions \cite{bea04,dal08} as a testing ground for saturation physics. A primary account of our approach was given in Ref. \cite{mtw09}. It made use of the color-glass framework \cite{mcl94,Bal96,Jal97,Ian01, nar05,alb07bis}. Related approaches of other authors to the net-baryon problem with respect to gluon saturation are Refs. \cite{ita03,alb07}. We  compare with data that were obtained by scaling distributions of identified net protons at Super Proton Synchrotron (SPS) and RHIC energies, and extrapolate to LHC energies. This problem was treated with different approaches based on QCD without saturation \cite{bas03,alv09,mi07}.

In \(A + A\) collisions, two distinct and symmetric peaks with respect to rapidity \(y\) occur at SPS energies \cite{app99} and beyond. The rapidity separation between the peaks increases with energy and decreases with increasing mass number, \(A\), reflecting larger baryon stopping for heavier nuclei, as was investigated phenomenologically in the nonequilibrium-statistical relativistic diffusion model \cite{wols06,wol06}. In this work we show how the evolution of the peaks can be linked to saturation physics.

The net-baryon number is essentially transported by valence quarks that probe the saturation regime in the target by multiple scatterings. During the collision, the fast valence quarks in one nucleus scatter in the other nucleus by exchanging soft gluons, leading to their redistribution in rapidity space. We take advantage of the fact that the valence quark parton distribution is well known at large \(x\), which corresponds to the forward and backward rapidity region, to access the gluon distribution at small \(x\) in the target nucleus. Therefore, this picture provides a clean probe of the unintegrated gluon distribution, \(\varphi(x,p_T)\), at small \(x\) in the saturation regime. Here \(p_T\) is the transverse momentum transfer.

In particular, we use net-baryon rapidity distributions in central relativistic heavy-ion collisions from SPS to LHC energies to probe saturation physics through their energy and mass-number dependence on a geometric scaling variable \(\tau\).

We discuss the net-baryon rapidity distributions from SPS to LHC energies in Sec. II. Their dependence on the scaling variable \(\tau\) is investigated and the position of the fragmentation peak as a function of the saturation-scale exponent \(\lambda\) is derived. Possible conclusions about saturation from the midrapidity valley in net baryons are drawn and the effect of quark fragmentation into hadrons is considered.

In Sec. III, theoretical results are compared with data on transverse momentum spectra for Au + Au at RHIC energies of 62.4  and 200 GeV and with rapidity spectra at SPS energies of 17.3 GeV, as well as at RHIC energies. Predictions for the rapidity distribution of net baryons at LHC energies are made. Net-kaon rapidity distributions are also discussed. Centrality and system-size dependence are investigated, and the mean rapidity loss as a function of beam rapidity, or energy, is discussed. Finally,  conclusions are drawn in Sec. IV.

\section{\label{sec:rap}Rapidity distributions in the color glass condensate}

The differential cross section for valence quark production in a high-energy nucleus-nucleus collision reads \cite{dum06,kha04}
\begin{equation} \label{eq:crossGS}
\frac{dN}{d^2p_Tdy}= \frac{1}{(2\pi)^2 } \frac{1}{ p_T^2}\;x_1q_v(x_{1},Q_{f})\;\varphi\left(x_2,p_T\right),
\end{equation}
where \(p_T\) is the transverse momentum of the produced quark and \(y\) its rapidity. The longitudinal momentum fractions carried, respectively, by the valence quark in the projectile and the soft gluon in the target are \(x_1=p_T/\sqrt{s}\exp(y)\) and \(x_2=p_T/\sqrt{s}\exp(-y)\). The contribution of valence quarks in the other beam nucleus is added incoherently by changing \(y \to -y\). The valence quark distribution of a nucleus, \(q_v\equiv q-\bar q\), is assumed to be given by the sum of valence quark distributions \(q_{v, N}\) of individual nucleons, \(q_v\equiv A q_{v, N}\), where \(A\) is the atomic mass number. 

The factorization scale is usually set equal to the transverse momentum, \(Q_{f}\equiv p_T\). Since the valence quark parton distribution is weakly dependent on \(Q_f\) we  neglect it in the following discussion. The gluon distribution is related to the forward dipole scattering amplitude \({\cal N}(x,r_T)\) \cite{dum06}, for a quark dipole of transverse size \(r_T\), through the Fourier transform
\begin{equation}\label{eq:FourierN}
\varphi(x,p_T)=2\pi p_T^2\int r_Tdr_T {\cal N}(x,r_T)J_0(r_Tp_T).
\end{equation}

Assuming the rapidity distribution for net-baryons is proportional to the valence quark rapidity distribution up to a constant factor of \(C\), we obtain, by integrating over \(p_T\),
\begin{equation} \label{eq:ydistquark}
\frac{dN}{dy}= \frac{C}{(2\pi)^2 }\int \frac{d^2p_T}{ p_T^2}\;x_1q_v(x_{1},Q_{f})\;\varphi\left(x_2,p_T\right).
\end{equation}
We  show that this is indeed a good approximation, at high energy, in the section devoted to the issue of fragmentation functions.

\begin{figure}
\includegraphics[width=7cm]{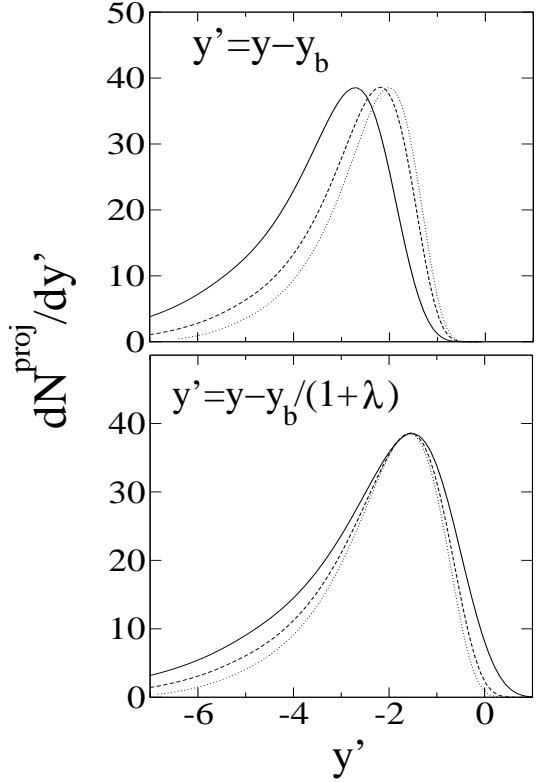}
\caption{\label{fig1} Projectile contributions to the net-baryon rapidity distribution plotted as a function of the variable \(y' = y-y_b\) (upper panel) for central Au + Au collisions at RHIC energies of \(\sqrt{s_{NN}} = \) 62.4 GeV (dotted curve) and 200 GeV (dashed curve) and for Pb + Pb at LHC energies of \(\sqrt{s_{NN}} =\) 5520 GeV (solid curve). The limiting fragmentation property is obviously broken for net baryons. The lower panel shows the equivalent curves for our variable \(y' = y - y_b/(1+\lambda)\) with \(\lambda =\) 0.2; here scaling is fulfilled in the peak region.}
\end{figure}

\subsection{Scaling property of the net-baryon distribution}\label{sec:geoscal}

Since the valence quark distribution is well known, net-baryon production represents a good observable to probe saturation effects in high-energy heavy-ion collisions through the respective gluon distributions. One important prediction of the color glass condensate theory is geometric scaling: The gluon distribution depends on \(x\) and \(p_T\) only through the scaling variable \(p_T^2/Q_s^2(x)\), where \(Q_s^2(x)=A^{1/3} Q_0^2\;x^{-\lambda}\); the scale \(Q_0\) sets the dimension. Geometric scaling was confirmed experimentally at HERA \cite{sta01}. The fit value \(\lambda = 0.2\)--0.3 agrees with theoretical estimates based on next-to-leading order Balitskii-Fadin-Kuraev-Lipatov (BFKL) results \cite{lip76,tri03}. To show that, in the high-energy limit, the net-baryon distribution reflects the geometric scaling of the gluon distribution, we perform the following change of variables:
\begin{equation}
x\equiv x_1,\;\;x_2\equiv x\;e^{-2y},\;\; p_T^2\equiv x^2 s\; e^{-2y}.
\end{equation}
Thus, we rewrite Eq. (\ref{eq:ydistquark}) as
\begin{equation}
\frac{dN}{dy}(\tau)=\frac{C}{2\pi}\int_0^1\frac{dx}{x}\;xq_v(x) \;\varphi(x^{2+\lambda}
e^\tau),
\label{eq:GSyield}
\end{equation}where \(\tau=\ln (s/Q_0^2) - \ln A^{1/3} - 2(1+\lambda)\,y\) is the corresponding scaling variable. Hence, the net-baryon multiplicity is only a function of a single scaling variable \(\tau\), which relates the energy dependence to the rapidity and mass-number dependence. From the equation for the isolines, \(\tau=\text{const}\), we obtain the evolution of the position of the fragmentation peak. This reflects the interplay of the valence quark distribution, peaked at \(x_1\sim 0.2\), with the gluon distribution peaked at \(p_T\sim Q_s\), in the forward region with respect to the variables of the problem
\begin{equation}\label{eq:peak}
y_{\text{peak}}=\frac{1}{1+\lambda}\left(y_{b}-\ln A^{1/6}\right)+\text{const},
\end{equation}
where \(y_b =(1/2) \ln[(E + p_{L})/(E - p_{L})]\simeq \ln(\sqrt s / m)\) is the beam rapidity at beam energy \(E\) and longitudinal momentum \(p_{L}\) with the nucleon mass \(m\).  

In Fig. \ref{fig1}\marginpar{F1}, we show the peak positions as given by Eq. (\ref{eq:peak}) for different incident energies. Hence, different values of \(y_b\), as functions of \(y'\) with \(y' = y - y_b\) in the upper panel, and \(y' = y - y_b/(1 + \lambda)\) in the lower panel, where \(y'\) is linearly related to our scaling variable \(\tau\) through \(\tau = - 2(1+\lambda)y'\) are shown. In this calculation we  include baryon mass effects by the replacement \(p_T\to \sqrt{p_T^2+m^2}\). We then observe clear violations of the \(y - y_b\) scaling in the upper panel. Hence, according to our model, limiting fragmentation phenomena as observed in particle production \cite{bac03}, where the rapidity distribution scales as \(y-y_b\), is violated in net-baryon rapidity distributions, which  should instead exhibit a scaling with \(y' = y-y_b/(1+\lambda)\)%\auq{Is this the correct meaning of the sentence beginning "Hence, according to our model, limiting fragmentation..."}. 
This is shown in the lower panel, where scaling is fulfilled in the peak region. The deviations outside the peak region are largely due to mass effects.
To draw this conclusion we do not yet need to specify the form of the gluon distribution, although for numerical calculations we must  specify one. The details of the computations are given in Sec. \ref{th-data}.

The compilation of the SPS data at \(\sqrt{s}=\)17.3 GeV \cite{app99} and RHIC at \(\sqrt{s}=62.4\) and 200 GeV \cite{dal08} provides an opportunity to verify the scaling law in Eq. (\ref{eq:peak}). The scaling properties of net-baryon rapidity distributions were investigated by the BRAHMS Collaboration for the one-nucleus contribution (see Ref. \cite{dal08} for more details about the subtraction procedure of the projectile contribution). Despite the lack of data in the peak region at RHIC, the data seem to exhibit a scaling with \(y-y_b\) compatible with the limiting fragmentation picture.

\begin{figure}
\includegraphics[width=7cm]{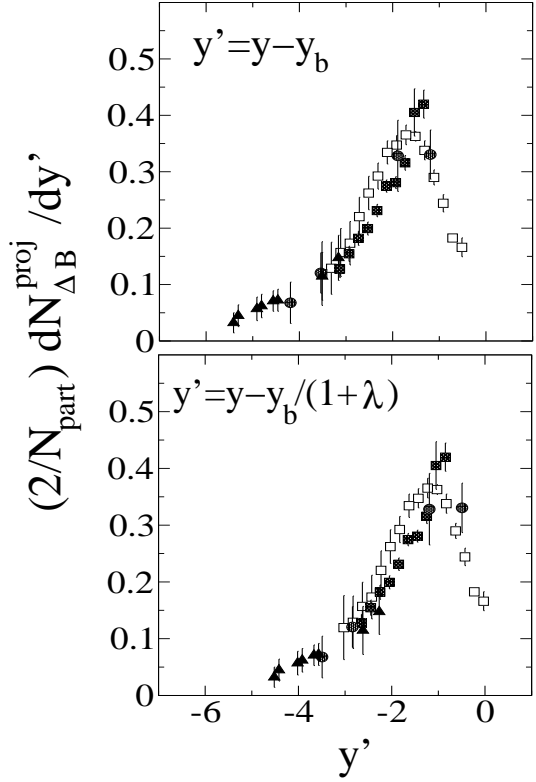}
\caption{\label{fig2} Measured projectile contributions to the net-baryon rapidity distribution plotted as a function of \(y' = y-y_b\) (upper panel) and \(y' = y - y_b/(1+\lambda)\) (lower panel), with a saturation-scale exponent \(\lambda =\) 0.2. Black squares are preliminary new Pb + Pb SPS data at \(\sqrt{s_{NN}} =\) 17.3 GeV \cite{blu08}, open squares are old Pb + Pb NA49 data \cite{app99}, circles are Au + Au BRAHMS data at 62.4 GeV \cite{dal08}, and triangles are BRAHMS data at 200 GeV \cite{bea04}. }
\end{figure}

A new and more refined analysis of SPS data was recently done  by the NA49 Collaboration \cite{blu08}, showing a slight discrepancy with the previous data \cite{app99}. The new data are shown as black squares in Fig. \ref{fig2}\marginpar{F2} for \(y' = y-y_b\) (upper panel) and for the scaling with \(y' = y-y_b/(1+\lambda)\) (lower panel) with \(\lambda=0.2\). The present data do not allow us to distinguish between the two different scaling laws. The value of the saturation-scale exponent \(\lambda=0.2\) was determined by recent calculations \cite{alb07} in the saturation picture, including running coupling effects, in agreement with particle production at RHIC. To improve the agreement with the new SPS data as compared to the analysis reported in Ref. \cite{mtw09}, which was based on the old NA49 data, we  increased \(\lambda\) from 0.15 to 0.2 in the present work. Note, however, that the applicability of the color glass condensate (CGC) picture may be questionable at the relatively low SPS energies.

In the peak region, the average \(x\) in the projectile is \(x\simeq 0.2\)--0.3, which corresponds to the average momentum fraction carried by a valence quark. In the target, \(x = (0.2\)--\(0.3)\,e^{-2y_{\text{peak}}}\), it decreases with increasing energy. In this kinematic regime we have a natural intrinsic hard momentum, the saturation scale \(Q_s\). This justifies the use of small-coupling techniques in QCD for calculating integrated yields \cite{dum03}. The effects of the medium are expected to be small at forward rapidity since the fast moving valence quarks escape the interaction zone quickly. A detailed measurement of the peak region will then enable us to reconstruct the gluon distribution from Eq. (\ref{eq:ydistquark}).

\subsection{Midrapidity valley in net protons}

It is worthwhile to analytically investigate  some limits of Eq. (\ref{eq:ydistquark}). In this section, we  derive a parametric formula for the region of small \(x_1\) (away from the peak), corresponding to the midrapidity valley (\(y\sim 0\)). In this kinematic regime the valence quark distribution behaves as \(xq_v\propto x^\Delta\), where \(\Delta\simeq 0.5\) is the intercept of the Regge trajectory \cite{ita03}, which allows us to perform analytic calculations.  

First, let us recall that the unintegrated gluon distribution is peaked at \(q_T=Q_s\), or \(x_1=\exp\left(-\tau/2+\lambda\right)\), reflecting the fact that most of the gluons sit at this value. Therefore, we  expect \(dN/dy\sim x_1q(\langle x_1\rangle)\), with \(\langle x_1\rangle\equiv \langle Q_s\rangle/\sqrt{s} \exp(y)\). Recalling that \(Q^2_s=A^{1/3} Q^2_0 x_2^{-\lambda}\) and \(x_2=x_1 \exp(-2y)\), we can solve the equation for \(\langle x_1 \rangle, \) yielding
\begin{equation}
\langle x_1 \rangle = \left(\frac{A^{1/6} Q_0}{\sqrt{s}}\right)^{1/(1+\frac{\lambda}{2})} \exp\left[2 \frac{1+\lambda}{2+\lambda} y\right]. \label{eq:meanx1}
\end{equation}
Finally, we obtain
\begin{equation}\label{eq:valley}
\frac{1}{A} \frac{dN}{dy} \propto \left(\frac{A^{1/6} Q_0}{ \sqrt{s} } \right)^{\Delta/(1+\frac{\lambda}{2})}\cosh\left[2\Delta \frac{1+\lambda}{2+\lambda} y\right].
\end{equation}

This property reflects the following asymptotic behavior at large \(\tau\), and \(x\ll 1\), of the gluon distribution:
\begin{equation}
\lim_{\tau\to\infty} \varphi(x,\tau)\propto\delta(x^{2+\lambda}e^\tau-1).
\end{equation}
This approximation is valid as long as \(\tau\) is large (and \(x\ll 1\)), typically in the midrapidity valley at asymptotic energies. For \(\lambda=0\) we recover Eq. (80) in Ref. \cite{ita03} plus the \(A\) dependence:
\begin{equation}\label{eq:valley0}
\frac{1}{A}\frac{dN}{dy}\propto \left(\frac{A^{1/6}}{\sqrt{s}}\right)^{\Delta} \cosh(2\Delta\;y).
\end{equation}
\subsection{The fragmentation function}

In this section we  investigate the effect of quark fragmentation on our conclusions about geometric scaling. By including the fragmentation function of a quark into net baryons, \(D(z) \equiv D_{\Delta B/q}(z)=D_{B/q}(z)-D_{\bar B/q}(z)\) (\(z=E_B/E_q\) being the fraction of the quark energy carried by the baryon fragment) and using isospin symmetry \(D(z) \equiv D_{\Delta B/q}(z)=-D_{\Delta B/\bar q}(z)\), the cross section for the production of a hadron of transverse momentum \(p_T\) at rapidity \(y\) reads
\begin{equation} \label{eq:crossFF}
\frac{dN}{d^2p_Tdy}= \frac{1}{(2\pi)^2 } \int_{x_{\text{F}}}^1 \frac{dz}{z^2} D(z)\frac{1}{q_T^2}\;x_1q_v(x_1)\;\varphi\left(x_2, q_T \right),
\end{equation}
where \(q_T=\sqrt{p_T^2+m^2}/z\) is the quark momentum, \(x_F=\sqrt{p_T^2+m^2}/\sqrt{s}\exp(y)\) is the Feynman-\(x\), and \(x_1=q_T/\sqrt{s}\exp(y)\), \(x_2=q_T/\sqrt{s}\exp(-y)\). Integrating Eq. (\ref{eq:crossFF}) over \(p_T\) up to the kinematic boundary \(p_{\text{max}}=\sqrt{s}\;e^{-y}\) imposed by \(x_F<1\), and inverting the order of the integrals, yields
\begin{eqnarray}
\frac{dN}{dy}&= & \int_0^{p_{\text{max}}}\frac{d^2p_T}{(2\pi)^2 }\int_{x_F}^1 \frac{dz}{z^2} D(z) \frac{1}{q_T^2}\;x_1q_v\left(x_1\right)\;\varphi\left(x_2, q_T\right)\nonumber\\
&=& \frac{1}{(2\pi)^2 } \int_{z_0}^1 dz D(z) \int_{m/z}^{q_{\text{max}}}\frac{d^2q_T}{q_T^2}\;x_1q_v(x_{1})\;\varphi\left(x_2, q_T \right),\nonumber\\\end{eqnarray}with \(z_0=m/\sqrt{s} \exp(y)\) and \(q_{\text{max}}=\sqrt{s} \exp(-y)\).  Now, if we assume geometric scaling we have
\begin{equation} \label{eq:GSFF1}
\frac{dN}{dy}= \frac{1}{2\pi } \int_{z_0}^1 dz D(z) \int_{z_0/z}^1 \frac{d x}{x}\; xq_v(x)\;\varphi\left(x^{2+\lambda}e^\tau\right).
\end{equation}
Obviously, since \(z_0\) depends on the energy and on the rapidity, it violates explicitly geometric scaling. However, in the high-energy limit, when \(s\rightarrow \infty\), or more precisely when \(m\ll\; \langle p_T\rangle \), the lower bound of the integral can be set to 0 and one recovers the geometric scaling formula of Eq. (\ref{eq:GSyield}) with \(C= \int_0^1 dz D(z) \).

At this stage, we  anticipate the discussion on the applicability of fragmentation functions for the observables of interest. The main contribution to the rapidity distribution comes from baryons of transverse momentum \(p_T \sim 1\) GeV, which is low enough to render the use of the fragmentation functions questionable. Moreover, it was pointed out clearly by Bass {and co-workers} \cite{fri03} that, for \(p_T\lesssim 5\) GeV, parton recombination dominates the hadronization process in particle production. Hence, the hadrons will be produced by partons of smaller energy instead of fragmentation of partons of larger energy as required by the fragmentation picture. Therefore, to simplify the discussion, we  assume \(D(z)\propto \delta(1-z)\) to account roughly for the competition between recombination and fragmentation.

To gain more insight into the hadronization process of the valence quarks, let us recall the relationship between the peak position and the parton transverse momentum. At the peak position, the peak of the valence quark distribution \(x_1\simeq 0.2\) is reflected. Hence, one can extract the mean transverse momentum as
\begin{equation}
\langle q_T\rangle\simeq 0.2 \sqrt{s} \exp(-y_{\text{peak}}).
\end{equation}
Moreover, the mean hadron transverse momentum \(\langle p_T\rangle\) can be extracted from the hadron spectra. Therefore, one obtains a measure of the energy that flows from the valence quark to the net protons (baryons) as
\begin{equation}
\langle z\rangle=\frac{\langle p_T\rangle}{\langle q_T\rangle}.
\end{equation}

For \(\langle z \rangle<1\), the hadronization process is dominated by fragmentation, in other words, the valence quarks lose energy by radiating gluons and, therefore, the produced baryon carries a fraction of the quark momentum. Furthermore, for \(\langle z \rangle>1\), the mean momentum of the hadron is larger than that of the valence parton, which can be achieved by parton recombination \cite{fri03}.

\section{Theory versus data} \label{th-data}

To take into account saturation effects in the target we choose the Golec-Biernat-W\"usthoff model \cite{GBW98} for the forward dipole scattering amplitude \(\cal N\), leading to (cf. Eq. (\ref{eq:FourierN}) and Ref. \cite{dum06})
\begin{equation}
\varphi(x,p_T)=4\pi\frac{p_T^2}{Q_s^2(x)}\exp\left(-\frac{p_T^2}{Q_s^2(x)}\right),
\end{equation}
in the fundamental representation of SU(3). This Gaussian form actually reflects  the multiple scatterings performed by the valence quarks in the color glass. It is interesting to make the connection to the fitting procedure performed by the BRAHMS Collaboration when integrating the spectra. Indeed, while the usual fitting functions used for particle production are exponential in \(p_T\) or have a Boltzmann form, the BRAHMS Collaboration noticed that, for net-proton spectra, the best fits are obtained with Gaussians in \(p_T\). This corroborates our picture.

First, we investigate transverse momentum distributions for net protons in comparison with BRAHMS data taken at different rapidities to fix \(Q_0\). The spectra constrain our model that contains only two parameters, \(Q_0\), and the overall normalization C. This investigation also provides some hints regarding the transition from a coherent gluonic state (the color glass condensate) at low transverse momenta to incoherent partonic (jetlike) interactions at high transverse momenta.

The valence quark parton distribution function (pdf) of the nucleus is taken to be equal to the valence quark pdf in a nucleon times the number of participants in the nucleus. Here, we  focus  on the forward rapidity region and interpolate to midrapidity where small-\(x\) quarks are dominant, by matching the leading-order distributions \cite{mrst01} and the Regge trajectory, \(xq_v\propto x^{0.5}\), at \(x =0.01\) \cite{ita03}.

We  compare the data to two calculations corresponding to different assumptions concerning hadronization of the valence quarks. First, we  use fragmentation functions for valence quarks to net-protons \cite{alb08} (the only ones that are presently available),
\begin{equation}
D_{p-\bar p}(z)=N\;z^a\;(1-z)^b,
\end{equation}
with \(N=520142\), \(a=11.6\), and \(b=6.74\) fitted at \(Q=1.4\) GeV. We can neglect QCD evolution in the range of interest. This is represented by dashed curves in Figs. \ref{fig3} and \ref{fig4}\marginpar{F3,4}. We  refer to this result as the F-model (for fragmentation model). In the NF-model (no-fragmentation), we  assume that, on average, the energy of the parton is equal to that of the produced baryon, namely \(D(z)\propto \delta(1-z)\). This assumption is based on the fact that hadronization at low \(p_T\) is poorly known and it was proven \cite{fri03} that it is dominated by recombination of partons instead of fragmentation, which will imply \(z>1\).

Hence, there should be a competition between the two phenomena, fragmentation and recombination. Namely, due to the high parton density characteristic of high-energy heavy-ion collisions, the parton showers described in the fragmentation picture can overlap. Hence, the recombination of partons from different showers can occur in addition to the recombination of shower partons with thermal partons \cite{fri03,hwa04}.

To fix the scale \(Q_0\) we  use the forward-rapidity net-proton spectra at 200 GeV where we expect our model to be valid: high energy and large rapidity. Whereas \(Q_0\) is fixed once for all the available energies, the normalization is tuned at each energy to fit  the data owing to our lack of knowledge on the hadronization process. Moreover, our calculation is at leading order, therefore, a K-factor is required.

For \(\lambda = 0.2\), we  fix \(Q_0^2=0.04\; \)GeV\(^2\) in the NF-model (\(0.1\; \)GeV\(^2\) in the F-model), leading to \(Q_s^2 = 0.6\) GeV\(^2\) (1.5 GeV\(^2\) in the F-model) at \(x = 0.01\). When comparing to investigations of charged-hadron production, such as in Refs. \cite{nar05,kha04}, which involve the gluon distribution in the adjoint representation of SU(3), we must consider a rescaling of our net-baryon \(Q_s^2\) by the color factor \(N_C/C_F\) with \(C_F = (N_C^2-1)/2N_C\) and \(N_C = 3\), corresponding to a factor 9/4.

\subsection{Transverse momentum spectra and rapidity distributions at 200 and 62.4 GeV}

\begin{figure*}
\includegraphics[width=17cm]{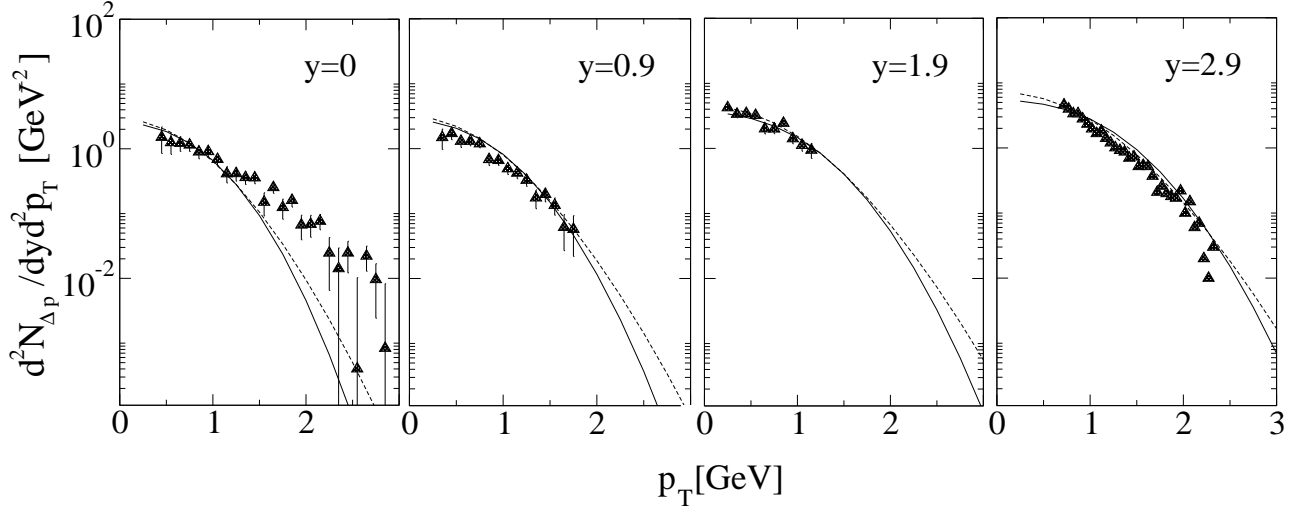}
\caption{\label{fig3} Net-proton spectra for Au + Au at RHIC energies of \(\sqrt{s_{NN}} = \) 200 GeV for different rapidities. From left to right, \(y=0, 0.9,\) 1.9, and 2.9. The triangles represent BRAHMS data \cite{bea04}; at \(y =2.9\), more recent preliminary BRAHMS data are shown \cite{chr09}. The saturation-scale exponent in the calculations is \(\lambda=\) 0.2. Full lines correspond to the no-fragmentation hypothesis with \(Q_0^2=0.04\; \)GeV\(^2\), whereas the dashed lines include fragmentation with \(Q_0^2=0.1\; \)GeV\(^2\).}
\end{figure*}

\begin{figure*}
\includegraphics[width=17cm]{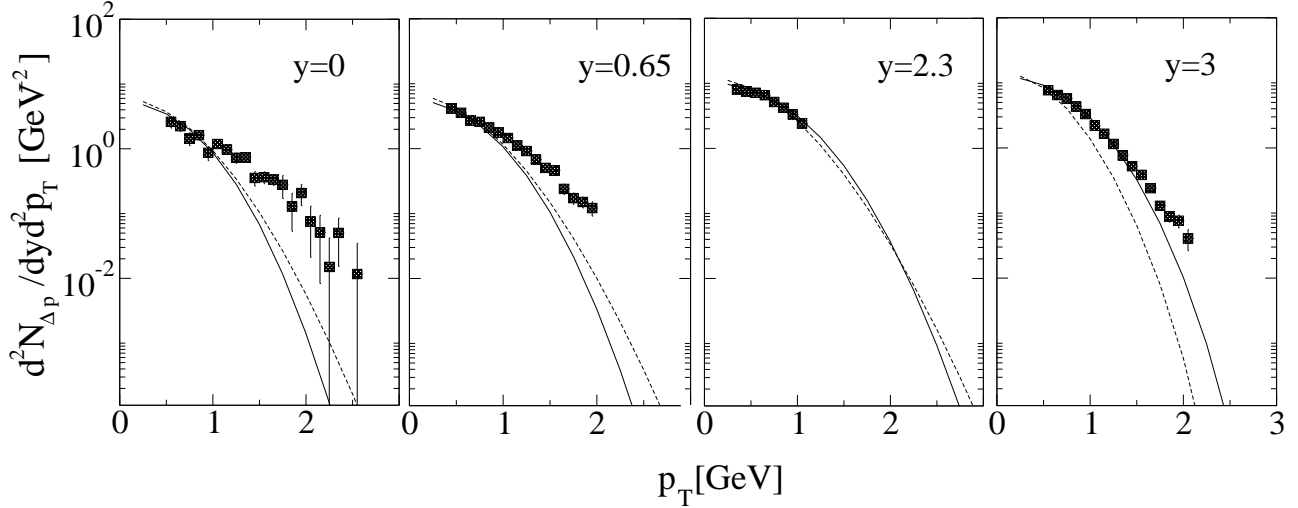}
\caption{\label{fig4} Net-proton spectra for Au + Au at RHIC energies of \(\sqrt{s_{NN}} = \) 62.4 GeV for different rapidities. From left to right, \(y=0,\) 0.65, 2.3, and 3. The squares represent the BRAHMS data \cite{dal08}. The full lines correspond to the no-fragmentation hypothesis with \(Q_0^2=0.04\; \)GeV\(^2\) and the dashed lines include fragmentation with \(Q_0^2=0.1\; \)GeV\(^2\). The saturation-scale exponent is \(\lambda=\) 0.2 in all cases. }
\end{figure*}

In Figs. \ref{fig3} and \ref{fig4}, the \(p_T\) spectra at 200  and 62.4 GeV are shown for a large rapidity range. Results for the NF-model (full curves) are compared with the F-model (dashed curves). The increasing discrepancy between the data and both models toward midrapidity is expected because of the reduction of the window for saturation effects with decreasing energy and rapidity. In addition, the quark-nucleus cross section is valid in the eikonal approximation: It is meant to be applicable at forward rapidity where the valence-quark longitudinal momentum fraction is \(x_1\sim1\). Another source of uncertainty can result from an additional rescattering of the valence quark in the quark-gluon plasma leading to an enhancement of the mean \(p_T\) broadening, essentially at midrapidity, where medium effects are expected to be important,
\begin{equation}
\langle p_T\rangle \simeq Q_s \to  \langle p_T\rangle\simeq Q_s+\langle p_T\rangle_{\text{med}},
\end{equation}
resulting in a harder spectrum at midrapidity and better agreement with the data.

\begin{figure*}
\includegraphics[width=17cm]{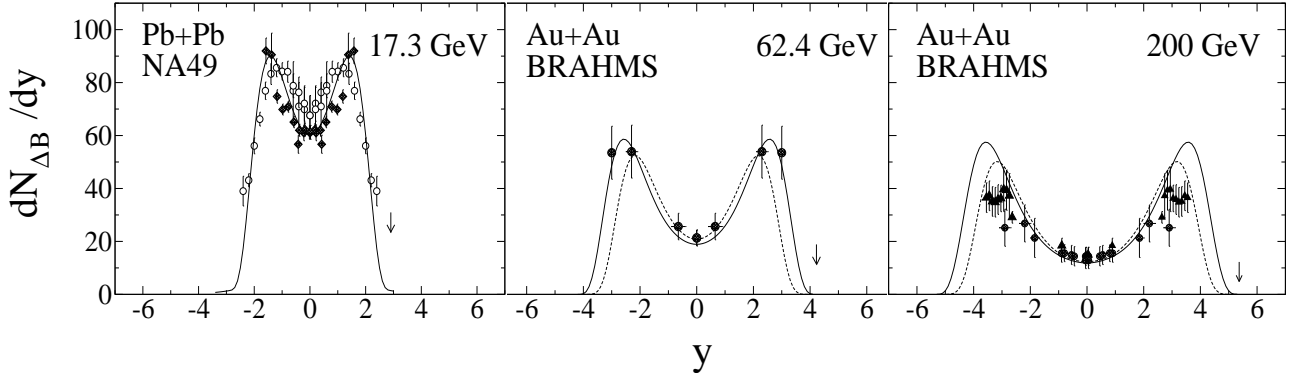}
\caption{\label{fig5} Rapidity distribution of net baryons in central (0\%--5\%) Pb + Pb collisions at SPS energies of \(\sqrt {s_{NN}}\) = 17.3 GeV (left frame). The theoretical calculation for \(\lambda =\) 0.2 and \(Q_0^2=0.04\; \)GeV\(^2\) is compared with NA49 results that were extrapolated from the net-proton data (open circles \cite{app99}). Black diamonds are more recent preliminary NA49 data points \cite{blu08}. At RHIC energies of \(\sqrt{ s_{NN}}\) = 62.4 GeV (middle frame, 0\%--10\%) and 200 GeV (right frame, 0\%--5\%) for central Au + Au, our corresponding theoretical results are compared with BRAHMS net baryon data (circles) \cite{bea04,dal08}. At 200 GeV, triangles are preliminary scaled BRAHMS net proton data points for 0\%--10\% \cite{deb08}. The full lines correspond to the no-fragmentation hypothesis with \(Q_0^2=0.04\; \)GeV\(^2\) and the dashed lines include fragmentation with \(Q_0^2=0.1\; \)GeV\(^2\). Arrows indicate the beam rapidities.}
\end{figure*}

The integrated net-proton rapidity distributions, scaled by a factor of 2.05 \cite{dal08} to obtain the net-baryon distributions, are shown in Fig. \ref{fig5}\marginpar{F5}. Whereas the NF-model (full line) describes the data well, the implementation of the fragmentation function gives an unsatisfactory result. The estimated numbers of participants are \(390\), \(315\), and \(357\) for \(\sqrt{s}=17.3\), 62.4, and 200 GeV, respectively \cite{dal08}. From the integration of the rapidity distribution of the NF-model about 15\% of the participant baryons are missing, most of them in the tails of the distributions, which are too steep in our high-energy model. Hence, the NF-model accounts for about 85\% of the estimated baryon number, whereas the F-model, which includes fragmentation, accounts for significantly fewer baryons. We believe that a better description of the hadronization process, in particular by incorporating recombination, will improve the overall agreement with the data.

\begin{figure}
\includegraphics[width=7cm]{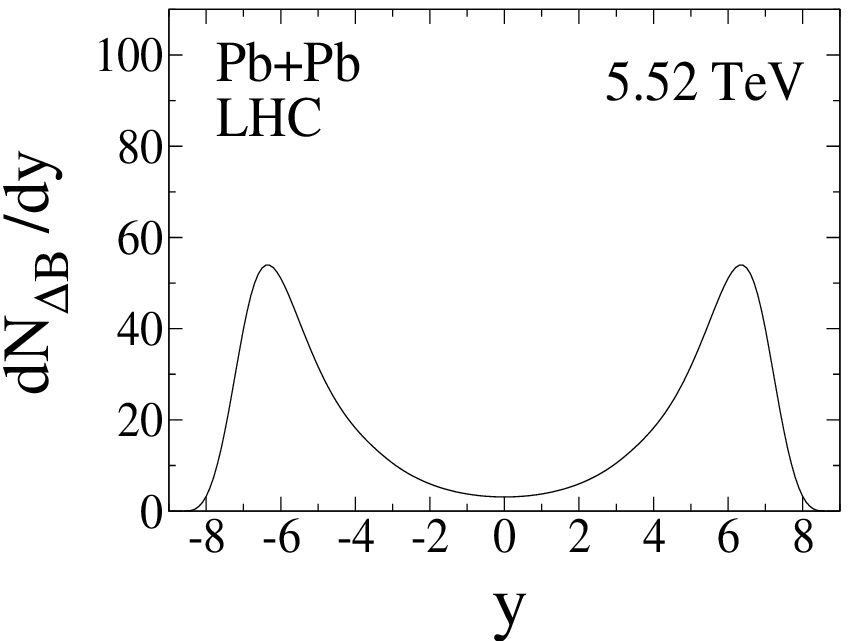}
\caption{\label{fig6}Rapidity distribution of net baryons in 0\%--5\% central Pb + Pb collisions at LHC energies of \(\sqrt {s_{NN}}\) = 5.52 TeV. The theoretical distribution is shown for \(\lambda =\) 0.2 and \(Q_0^2=0.04\; \)GeV\(^2\). }
\end{figure}

We  extrapolate our results for the NF-model to central (0\%--5\%) Pb + Pb collisions at LHC energies of 5.52 TeV in Fig. \ref{fig6}\marginpar{F6}, where the theoretical distribution is shown for \(\lambda =\) 0.2 and \(Q_0^2=0.04\; \)GeV\(^2\). Since \textsc{alice}  measures identified protons and antiprotons only in the rapidity range \(|y|<2\), a direct observation of the fragmentation peaks, and of the dependence of the peak position on the saturation-scale exponent as discussed in Ref. \cite{mtw09}, will depend on future extensions of the LHC heavy-ion detectors to enhance forward capabilities.

A further test of our model, and of the value of the deduced saturation scale, is provided by comparing it with net-kaon distribution functions and, in particular, \(dN/dy\) for net kaons. The bulk of the produced \(K^{+}\) and \(K^{-}\) mesons are due to inclusive gluon production in the midrapidity source. The net kaon distribution ( \(K^{+} - K^{-}\)), however, is essentially due to the interaction of fast valence quarks with soft gluons in the target, just as in the case of the net-proton distribution that we  have discussed thus far.

\begin{figure}
\includegraphics[width=7cm]{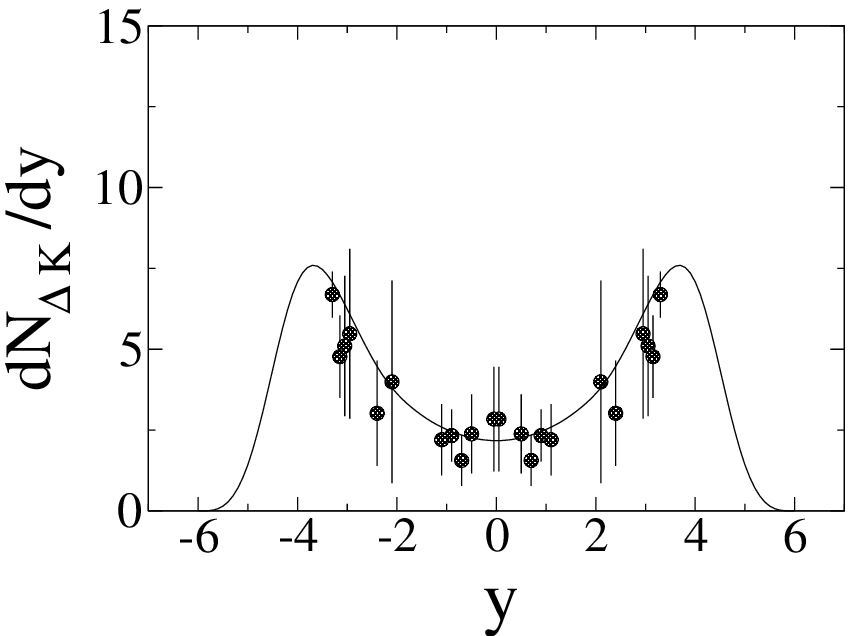}
\caption{\label{fig7}Net-kaon rapidity distribution as obtained from our model in comparison with the BRAHMS data \cite{bea05} for central (0\%--5\%) Au + Au at \(\sqrt {s_{NN}}\) = 200 GeV. The calculation is for \(\lambda = 0.2\) and \(Q_0^2=0.04\) GeV\(^2\). }
\end{figure}

The net-kaon distribution that is obtained from the 200-GeV BRAHMS data is shown in Fig. \ref{fig7}\marginpar{F7}. The data show the same trend as the net-baryon results. For kaons, the net-charge content is 44. (In the total charge balance of the collision, we must consider all particle species. Since negative pions are slightly more abundant than positive pions, the integral of net-proton and kaon distributions may actually exceed 158.) The calculation, as shown in Fig. \ref{fig7}, agrees well with these data.

\subsection{\label{sec:centrality} Centrality and system-size dependence}

We  also investigate the centrality dependence of the net-baryon distribution. Formally, in Sec. \ref{sec:rap}, we  show the dependence of the rapidity distribution on the mass number \(A\) through the saturation scale \(Q_s\propto A^{1/6}\). The centrality dependence of particle production is essentially determined by the number of participants and hence we now make the replacement \(A\to N_{\text{part}}\).

\begin{figure}
\includegraphics[width=7cm]{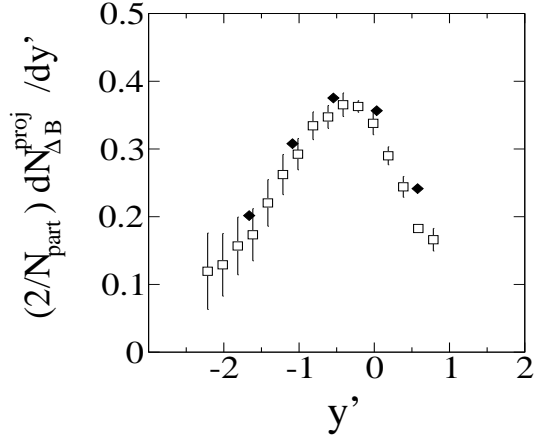}
\caption{\label{fig8} Pb + Pb net-baryon rapidity distribution at \(\sqrt {s_{NN}}\) = 17.3 GeV (open squares are NA49 data \cite{app99}) and S + S rapidity distribution at \(\sqrt {s_{NN}}\) = 19.4 GeV (black diamonds are NA35 data \cite{alb98}) plotted as functions of the scaling variable \(y'=y-(y_b-\ln N_\text{{part}}^{1/6})/(1+\lambda)\) with \(N_\text{{part}}\) = 352 for Pb + Pb and \(N_\text{{part}}\) = 52 for S + S.}
\end{figure}

Let us recall first a general feature of net-baryon distributions as a function of system size \(A\): More stopping is observed for larger systems. In other words, it corresponds to the shift of the fragmentation peaks 
%\revise{towards}{toward} 
toward midrapidity with increasing \(A\). To quantify this,  we consider two experiments involving different nuclei, \(A_1\) and \(A_2\) such that \(A_1>A_2\gg1\), then according to Eq. (\ref{eq:peak}) for the net-baryon peak position, we obtain for the rapidity difference of the peaks for two different mass numbers:
\begin{equation}\label{eq:cent-peak}
\Delta y_{\text{peak}}=y_{\text{peak}}(A_1)-y_{\text{peak}}(A_2)=-\frac{1}{1+\lambda} \ln \left(\frac{A_1}{A_2}\right)^{1/6},
\end{equation}
where the negative sign reflects the increasing stopping with increasing \(A\). Let us consider a central gold-gold collision with \(N_{\text{part}}\simeq 350\equiv A_1\) and a peripheral one with \(N_{\text{part}}\simeq 50\equiv A_2\). Using Eq. (\ref{eq:cent-peak}), for \(\lambda=0.2\), we obtain for the rapidity shift \(\Delta y_{\text{peak}}\simeq 0.27\). Hence, provided the measurements of the peak region are precise, we  access  the value of \(\lambda\) by measuring the peak shift. Unfortunately, the data are not completely conclusive yet. In Fig. \ref{fig8}\marginpar{F8,9}, we show that  the data for Pb + Pb collisions at 17.3 GeV with \(N_\text{{part}} = 352\) in the measured rapidity range (390 if one extrapolates to full rapidity range) indeed scales with the data for S + S collisions with \(N_\text{{part}} = 52\) in the measured region \cite{alb98} according to Eq. (\ref{eq:peak}).

\begin{figure}
\includegraphics[width=7cm]{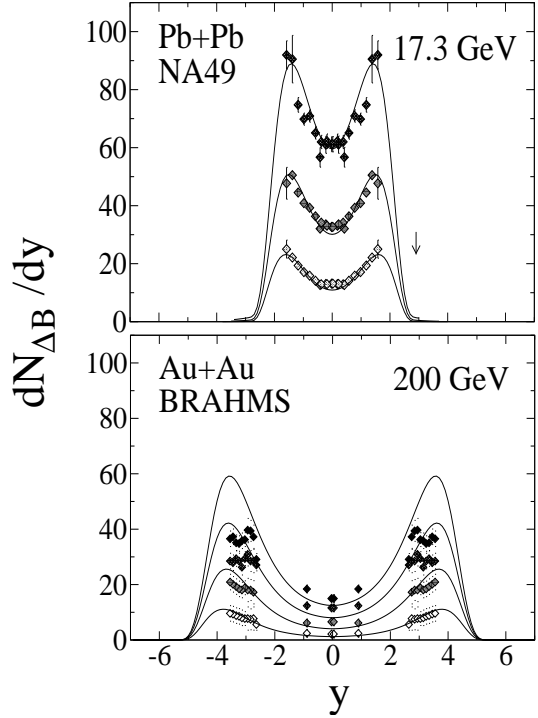}
\caption{\label{fig9} Rapidity distribution of net baryons in Pb + Pb collisions at SPS energies of \(\sqrt {s_{NN}}\) = 17.3 GeV (upper frame). The theoretical calculations for \(\lambda =\) 0.2 and \(Q_0^2=0.04\; \)GeV\(^2\) are compared with recent preliminary NA49 results \cite{blu08} for centralities of 0\%--5 \%, 12.5\%--23.5\%, and 33.5\%--43.5\% (top to bottom; net-proton data scaled to net baryons). At RHIC energies of \(\sqrt{ s_{NN}}\) = 200 GeV (bottom frame) for Au + Au, our corresponding theoretical results without fragmentation are compared with preliminary BRAHMS net baryon data for centralities of 0\%--10\%, 10\%--20\%, 20\%--40\%, and 40\%--60\% \cite{deb08}. }
\end{figure}

We show in Fig. \ref{fig9} the computation resulting from the NF-model,  the centrality dependence of the rapidity distribution at 17.3 and 200 GeV. Whereas the 17.3-GeV and the more peripheral 200-GeV theoretical curves agree with the data, the more central 200-GeV curve does not reproduce the absolute magnitude of the data at forward rapidity (see also Fig. \ref{fig5}).

\subsection{\label{sec:raploss}Mean rapidity loss}

With increasing energy the peaks move apart, and the solutions behave like
% \revise{travelling}{traveling} 
 traveling waves in rapidity space \cite{mun03}, which can be probed experimentally at distinct values of the beam energy or the corresponding beam rapidity. We  derive the peak position as a function of the beam rapidity as \(y_{\text{peak}} = v\, y_{ b} + \text{const}\) with the peak velocity \(v = 1/(1+\lambda)\) [cf. Eq. (\ref{eq:peak})]. The position of the peak in rapidity space as a function of the beam energy can, in principle, be determined experimentally or at least estimated (RHIC). Theoretically, its evolution with energy provides a measure of the saturation-scale exponent \(\lambda\). Hence, a precise determination of the net-baryon fragmentation peak position as a function of beam energy will provide detailed information about the gluon saturation scale.

\begin{figure}
\includegraphics[width=8.6cm]{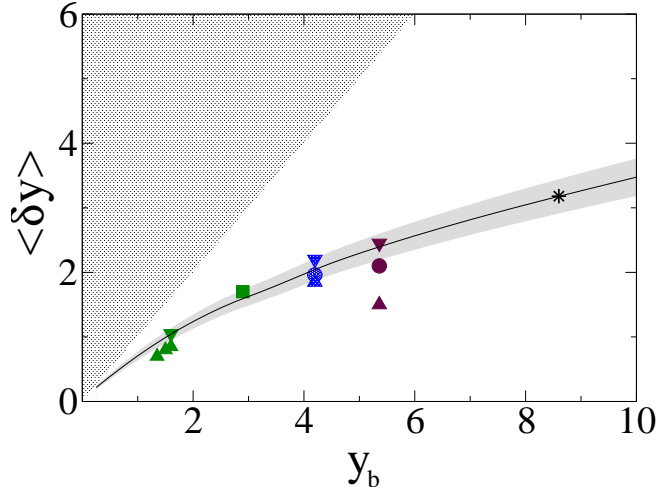}
\caption{\label{fig10} (Color online). The mean rapidity loss \(\langle \delta y\rangle\) as obtained from our theoretical results is plotted as a function of beam rapidity \(y_{b}\) (solid curve). The star at \(y_{\text{beam}}\) = 8.68 is our prediction for central Pb + Pb at LHC energies of \(\sqrt {s_{NN}}\) = 5.52 TeV with \(\lambda = 0.2\) and \(Q_0^2=0.04\) GeV\(^2\). Analysis results from AGS Au + Au data (E917, E802/E866, triangles) \cite{vi95}, SPS Pb + Pb data (NA49, square) \cite{app99}, and RHIC Au + Au data (BRAHMS, dots, with triangles as lower and upper limits) \cite{bea04,dal08} are compared with the calculations.}
\end{figure}

The mean rapidity loss \(\langle \delta y\rangle=y_{b}-\langle y\rangle\) is shown in Fig. \ref{fig10}\marginpar{F10}. Our result is in agreement with the experimental values of baryon stopping that was obtained at Alternating Gradient Synchrotron (AGS) and SPS energies \cite{vi95,app99}. Assuming that the mean rapidity evolves similarly to the peak position, \(\langle y\rangle\equiv y_{\text{peak}} \)+ const, we show that the mean rapidity loss increases linearly at large \(y_{b}\):
\begin{equation}
\langle \delta y\rangle = \frac{\lambda}{1+\lambda}y_b+\text{const},
\end{equation}\\
where the slope is related to \(\lambda\). Hence, the mean rapidity loss that accompanies the energy loss in the course of the slowing down of baryons provides a measure for \(\lambda\) and thus a test for saturation physics. The gray band in Fig. \ref{fig10} reflects the uncertainty of how to place the remaining baryons that are missing in our model. We  refer to them by \(\Delta N\), with \(N\equiv N_{\text{part}}\). The upper limit corresponds to the case where the missing baryons sit at the mean rapidity, roughly about the peak rapidity. Then the corrected mean rapidity loss is equivalent to the theoretical one, \(\langle \delta y\rangle_{\text{corr}}\equiv \langle \delta y\rangle\). The lower limit corresponds to the case where they sit at the beam rapidity, \(\langle \delta y\rangle_{\text{corr}}\equiv (1-\Delta N/N) \langle \delta y\rangle\). The full line is the mean value of the two calculations and it is in reasonable agreement with the upper limit of the data given by BRAHMS.

\section{\label{sec:concl}Conclusion}

In summary, we  present a saturation model for net-baryon distributions that successfully describes net-proton rapidity distributions and their energy and mass dependence. The remarkable feature of geometric scaling predicted by the CGC is reflected in the net-baryon rapidity distribution, providing a direct test of saturation physics.

In particular, we  show that the peak position in net-proton or net-baryon rapidity distributions of centrally colliding heavy ions at ultrarelativistic energies obeys a scaling law involving the mass number and the beam energy. We  explore the validity of that scaling law in comparisons with experimental rapidity distributions from central Pb + Pb and Au + Au collisions at SPS and RHIC energies and find good agreement with the NA49 and BRAHMS data on net-proton distributions that were scaled to net baryons.

We  simultaneously investigate net-proton transverse momentum spectra at RHIC energies, as  measured by the BRAHMS Collaboration, to find reasonable agreement of our model results with the data away from midrapidity. There is an increasing discrepancy toward midrapidity, which is expected because the window for saturation effects is reduced with decreasing energy and rapidity.

The parameters saturation-scale exponent \(\lambda\) and momentum scale \(Q_0^2\) are determined from the transverse momentum spectra for Au + Au at 200 GeV and then kept fixed at all energies (SPS, RHIC, and LHC) in the calculations for the \(p_T\) spectrum at 62.4 GeV, and in all calculations of rapidity distributions. The third parameter is the overall normalization constant C, which we adjust to the data.

Model calculations for the rapidity distribution in central Pb + Pb collisions at LHC energies of 5.52 TeV are obtained; we previously discussed the dependence of the position of the fragmentation peak on the gluon saturation-scale exponent \(\lambda\)  in Ref. \cite{mtw09} and do not repeat it here.

Our analytical scaling law yields an excellent description of the mass dependence of the net-baryon distribution at SPS energies in a comparison of S + S and Pb + Pb results. The centrality dependence of the Pb + Pb rapidity distribution at 17.3 GeV is well reproduced, whereas discrepancies remain for Au + Au at RHIC energies.

The theoretical result for the mean rapidity loss in \(\sqrt{ s_{NN}}\) = 200 GeV Au + Au is larger than the BRAHMS result as derived from their data, but consistent with the experimental upper limit. This emphasizes the importance of a detailed analysis at LHC energies, where it may then be possible to determine the value of the saturation-scale exponent \(\lambda\) more accurately \cite{mtw09} and establish the attainment of gluon saturation.

To achieve this, forward measurements of identified hadrons---and in particular, baryons---for central heavy-ion collisions will be desirable. The ATLAS and Compact Muon Solenoid (CMS) detectors for \(p + p\) collisions are being extended to the forward region already; TOTEM is a dedicated forward detector. It will be useful to exploit these capabilities for central Pb + Pb physics as well.
%\auq{CMS spelled out; correct?}
\section*{ACKNOWLEDGMENTS}
This work was supported
by the Deutsche Forschungsgemeinschaft under Grant No. STA 509/1-1 and the ExtreMe Matter Institute, EMMI.

%\auq{Some references were reordered to cite in proper sequence. Refs. 42 and 43 need to be cited (42 $was originally uncited 29 and 43 was uncited 41)}

\end{document}